\documentclass[12pt,leqno]{article}
\usepackage{amsmath,amsthm,amsfonts,amssymb,color,mathrsfs,framed}
\usepackage{stmaryrd}

\DeclareMathAlphabet{\mathpzc}{OT1}{pzc}{m}{it}

\newtheorem{thm}{Theorem}[section]
\newtheorem{cor}[thm]{Corollary}

\theoremstyle{definition}
\newtheorem{defn}{Definition}[section]

\theoremstyle{remark}
\newtheorem{rem}{Remark}[section]

\definecolor{wco}{rgb}{0.5,0.2,0.3}

\numberwithin{equation}{section}

\begin{document}

\title{Noether theorems and quantum anomalies}
\author{John Gough$^{1}$, Tudor S. Ratiu$^{2}$, Oleg G. Smolyanov$^{3}$}
\addtocounter{footnote}{1}

\footnotetext{Institute of Mathematics, Physics and Computer
Sciences, Aberystwyth University, Great Britain}
\addtocounter{footnote}{1}
\footnotetext{School of Mathematics,
Shanghai Jiao Tong University, 800 Dongchuan Road, Minhang
District, Shanghai, 200240 China and Section de Math\'ematiques,
\'Ecole Polytechnique F\'ed\'erale de Lausanne, CH--1015 Lausanne,
Switzerland. Partially supported by Swiss NSF grant SwissMAP.
\texttt{ratiu@sjtu.edu.cn, tudor.ratiu@epfl.ch}
\addtocounter{footnote}{1} } \footnotetext{Faculty of Mechanics
and Mathematics, Lomonosov Moscow State University, Moscow,
Russia. Partially supported by the Russian Fund for Fundamental
Research. \texttt{smolyanov@yandex.ru} \addtocounter{footnote}{1}
}
\maketitle
\date{}
\begin{abstract}
In this communication, we show that both infinite-dimensional versions
of Noether's theorems, and the explanation of quantum anomalies can be
obtained using similar formulas for the derivatives of functions whose
values are measures (\cite{SmvW1995}) or pseudomeasures (\cite{GoRaSm2015}).
In particular, we improve some results in \cite{GoRaSm2015}.
\end{abstract}

\allowdisplaybreaks

\section{Introduction}

A quantum anomaly is the violation of the symmetry with respect to a given
transformation during a quantization procedure. As such, a
quantum anomaly occurs if a quantization procedure of a classical system,
invariant relative to a transformation, yields a quantum system which is no
longer invariant under this transformation. The calculations that are
used to explain this phenomenon parallel those appearing in the proof
of the infinite dimensional versions of Noether's theorems
(cf. \cite{SmvW1998}). Both these
calculations have two versions. In the first version, one uses
integration with respect to a Feynman pseudomeasure in the case of a
quantum anomaly, and alternatively integration with respect to a usual smooth
$\sigma$-additive measure in the case of Noether's theorem. In the
second version, both for quantum anomalies and Noether's theorem, one
employs integration with respect to a Lebesgue pseudomeasure. It is worth
noticing that the difference between Noether 's theorem and quantum
anomalies is only that in the calculations related
to the Noether theorem, one integrates the Lagrangian of the system (to get
the action), whereas in the calculations related to quantum anomalies, one
integrates the complex exponent of the action or of a part of the action.
Using these calculations, we analyze a famous contradiction between the
points of view presented in the well-known monographs \cite{CWM} and
\cite{FS}. In particular, in \cite[page 352]{CWM}, it is claimed
that the explanation of the origin of quantum anomalies given in \cite{FS}
is not correct. This criticism refers to the first, 2004 edition, of
\cite{FS}. The second, 2013 edition of \cite{FS}, does not address this
criticism, the authors maintain their original claims, and they do not
even quote \cite{CWM}. Using a mathematically rigorous alternative approach,
we arrive to the conclusion presented in \cite{FS}. It is worth mentioning
that our approach, being infinitesimal (like in the Noether
theorems), is conceptually much simpler than the global (and non-rigorous)
approach presented in \cite{CWM} and \cite{FS}. We also mention that in
our approach to both the Noether theorem and to the quantum anomalies,
we need not assume that the considered transformations are elements of
any group of symmetries of the system and hence Klein's Erlangen Program
has no direct relation to the subject of the present communication (already
in the classical book \cite{GeFo1963}, the authors do not mention any
group structure on the set of transformations).

The paper is organized as follows. In section \ref{sec:derivation}, we introduce some
definitions of derivatives of (pseudo)measure-valued functions; most of
these are more or less standard. In section \ref{sec:Noether}, we present some
calculations of derivatives of some functions whose values are products of
a measure or pseudomeasure on a locally convex space (LCS) $E$,
with a function depending on $x \in E$ and on the values at $x$ taken by
another function $g$, defined on $E$, and its derivative at $x$ (the
generalization to higher derivatives is straightforward) and make some
remarks about how these calculations can be used in Noether-type theorems.
In section \ref{sec:QA}, we get, as a corollary to calculations of the
preceding section, an explanation of the origin of quantum anomalies.
Like in our preceding paper \cite{GoRaSm2015}, we consider here instead
of integrals, some measures and pseudomeasures, which simplifies the
situation.

We discuss here only the algebraic structure of these problems and not
formulate precise analytical assumptions.

\section{Derivatives of (pseudo-)measure valued functions}
\label{sec:derivation}
The following definitions are more or less standard. For any LCS $E$,
let $\mathfrak{M}(E)$ be the vector space of all
(signed) Borel $\sigma$-additive measures on $E$, equipped with the weak
topology defined by a vector space $C_b(E)$ of some bounded functions
in natural duality with $\mathfrak{M}(E)$. A function $f:(a,b) \rightarrow
\mathfrak{M}(E)$ is said to be \emph{differentiable at} $t \in (a,b)$
if the limit $f'(t): = \lim_{\Delta t \to 0} (\Delta t)^{-1}
\left(f(t+ \Delta t) - f(t) \right)$, called the \emph{derivative of
$f$ at $t$}, exists and
if the measure $f'(t)$ is absolutely continuous with respect to $f(t)$,
denoted by $f'(t) \ll f(t)$. In this case, the Radon-Nikodym density
of $f'(t)$ with respect to $f(t)$ is called the \emph{logarithmic
derivative of $f$ at $t$} and is denoted by $\beta_f(t)$; thus, $f'(t)
= \beta_f(t) f(t)$.

Completely similar definitions can be also formulated
for functions taking values in a space of distributions
(pseudomeasures) that is defined as the space $(\mathscr{D}(E))'$ of all
continuous linear functionals
on a space $\mathscr{D}(E)$ of some real or complex valued differentiable
functions on $E$. We assume that $(\mathscr{D}(E))'$ is equipped with a
locally convex topology; the space $\mathscr{D}(E)$ need not coincide
with the usual Schwartz space of test functions, even if $E= \mathbb{R}^n$.
If a function $f: (a,b) \rightarrow (\mathscr{D}(E))'$ is differentiable and
if there exists a function $\beta_f(t)$ on $E$, which is a multiplier
in $(\mathscr{D}(E))'$ and such that $f'(t) = \beta_f(t)f(t)$, then
$\beta_f(t)$ is again called the \emph{logarithmic derivative} of
$f_{\nu, k}$ at $t$.

From now, $E$ always denotes an LCS.

\begin{defn}
Let $k$ be a vector field on $E$, i.e., $k$ is a Borel map of $E$ into
itself, $\nu \in \mathfrak{M}(E)$, $\varepsilon>0$. Define $S_k(t):=
x - tk(x)$ for $t \in (-\varepsilon, \varepsilon)$, $x\in E$, and
$f_{\nu, k}:(-\varepsilon, \varepsilon) \rightarrow \mathfrak{M}(E)$ by
$f_{\nu, k}(t):=(S_k(t))_* \nu$. The measure $ \nu$ is called
\emph{differentiable along the vector field} $k$ if the function \
$f_{\nu, k}$ is differentiable at $0$; then $f_{\nu, k}'(0)$ is called the
\emph{derivative of $\nu$ along $k$} and is denoted by $\nu_k'$; in
this case, the logarithmic derivative of $f_{\nu, k}$ at $t=0$ is called the
\emph{logarithmic derivative of $\nu $ along $k$} and is denoted by
$\beta^\nu_k$ (so $\beta^\nu_k$ is an almost everywhere defined function
on $E$). If $h \in E$ and $k _h(x): = h$ for all $x \in E$, then the measure $\nu $ is said to be differentiable along (the vector) $h$ if it is
differentiable along $k_h$; in this case, the
logarithmic derivative of $\nu $ along $k_h$ is called the \emph{logarithmic derivative of $\nu $ along $h$} and is denote by
$\beta^\nu (h ,\cdot )$, i.e., $\beta^\nu (h,x):= \beta^\nu_{k_h}(x) $
a.e.
\end{defn}

\begin{defn}
A vector subspace $H \subset E$ is called a \emph{locally convex} (in
particular, a \emph{Banach} or \emph{Hilbert}) \emph{subspace} of $E$, if
$H$ is endowed with the structure of an LCS (respectively, Banach, or Hilbert space), with respect to which the canonical embedding
$H \hookrightarrow E$ is continuous.
\end{defn}

If $H$ is a locally convex subspace of $E$ and $\nu \in \mathfrak{M}(E)$
is differentiable along each $h \in H$, then it can be shown (cf.
\cite{GoRaSm2015}) that both mappings $H\ni h \mapsto \nu'h \in
\mathfrak{M}(E)$ and $H\ni h \mapsto \beta^\nu(h, \cdot)$ are linear.
So $\nu$ is differentiable along each $h \in H$ if and only if the
mapping $\psi_\nu: H\ni h \mapsto \nu_h \in \mathfrak{M}(E)$, where
$\nu_h(A):=\nu(A+h)$ for any Borel set $A \subset E$, is G\^ateau
differentiable  and for any $h, k \in H$, the measure $(\psi_\nu)'(k)(h)$
is absolutely continuous with respect to $\nu_k$. Then the mapping
$E\ni x \mapsto \left[H\ni k \mapsto \beta^\nu(k, x)\right] \in H'$ is
called the \emph{logarithmic derivative of $\nu$ along $H$} (if $H$ is
a Hilbert space, $H'$ is identified with $H$; in this case, the mapping
$\overline{\beta}^\nu: E \rightarrow H$, defied by
$(\overline{\beta}^\nu(x), k): = \beta^\nu(k, x)$, is called the
\emph{logarithmic gradient}, but we will not use this terminology).

The notion of differentiability of a measure $\nu\in \mathfrak{M}(E)$
along a locally convex subspace $H \subset E$ can be generalized in
the following way. Let $H$ be a LCS, which need not be a subspace of $E$.

\begin{defn}
A map $\psi: H
\rightarrow \mathfrak{M}(E)$ is called differentiable at $h \in H$, if
it is G\^ateau differentiable at $h$ and if, for any $k \in H$, the
measure $\psi'(h)k$ is absolutely continuous with respect to $\psi(h)$.
The Radon-Nikodym derivative of $\psi'(h)k$ with respect to $\psi(h)$ is
called the \emph{logarithmic derivative of $\psi$ at $h$
along $k$} and is denoted by $\beta^\nu_\psi(h)k$.
The linear mapping $k \mapsto \beta^\nu_\psi(h)k$ of $H$ into a space of
functions on $E$ is called the \emph{logarithmic derivative of $\psi$ at
$h \in H$}.
\end{defn}

\begin{rem}
If $H$ is a locally convex subspace of $E$,  $\nu$ has a logarithmic
derivative along $H$,  and $k$ is a vector field in $E$, such that
$k(E) \subset H$, then the function $x \mapsto \beta^\nu(k(x), x)$ is,
in general, not well defined, because for $h \in H$ the function
$\beta^\nu (h, \cdot )$ on $E$ is defined only $\nu$-almost everywhere,
the domain being dependent on $h$. A possible definition of
$\beta^\nu(k(\cdot) , \cdot )$ is given in \cite{SmvW1998}, correcting
an older definition of the same function presented in \cite{SmvW1995}.
We recall the formula
\begin{equation}
\label{rem_formula}
\beta^\nu_k(x) = \beta^\nu(k(x),x) + \operatorname{tr}k'(x)
\end{equation}
from \cite{SmvW1995}, which will be used below.
\hfill $\lozenge$
\end{rem}

\section{Noether type theorems}
\label{sec:Noether}
For any locally convex space $G$ and any locally convex subspace
$H \subset E$, let $C_H(E,G)$ be the space of all mappings of
$E$ into $G$ that are infinitely differentiable along $H$ (for the definition, see \cite{GoRaSm2015}).

Let $G, Z$ be LCS, $H$ a Hilbert subspace of $E$, and $\nu \in
\mathfrak{M}(E)$. For any $g \in C_H(E, G)$, let $\mu(g, \nu)$ be the
measure on $E$ defined by $\mu(g, \nu):=
L(\cdot , g(\cdot ), g'(\cdot )) \nu$, where $L\in C_{\mathscr{H}}
(E \times G \times \mathscr{L}(H,G), \mathbb{C})$ and $\mathscr{H}:=
H \times G \times \mathscr{L}(H,G)$ ($L$ can be considered as a generalization of a Lagrange function).

Let $F: Z \times E \times G \times \mathscr{L}(H, G) \rightarrow E \times G$
be an infinitely differentiable mapping satisfying $F(0, x,r, \alpha) =
(x, r)$ for any $\alpha \in \mathscr{L}(H,G)$. One can think that $F$
defines a family of transformations of $E \times G$ depending on a parameter
$z \in Z$ and also on elements of $\mathscr{L}(H, G)$ that are, in
applications, the values of the derivative of a mapping from $E$ to $G$.

Define the associated
mapping $F_Z$ on $Z$ with values in the
space of mappings from $E \times G \times \mathscr{L}(E, G)$ to $E \times G$
by $F_Z(z)(x_1, x_2, \alpha) : = F(z, x_1, x_2, \alpha) \in E\times G$.

For a mapping $g \in C_H(E,G)$ and $\nu\in \mathfrak{M}(E)$, the mapping
$F_Z$ defines the measure-valued function $F_{g, \nu}:Z \rightarrow
\mathfrak{M}(E)$ in two steps. First, a $G$-valued function $g_Z$ is
defined by its graph $\{F_Z(z)(x, g(x), g'(x)) \mid x \in E\} \subset
E \times G$, assuming that this set is the graph of a function.
The function $g_Z$ can also be defined by
\[
g_Z(x_z): = F_{Z,2}(z) \left(g\left(\left(F_{Z,1}(z)\right)^{-1}(x_z)
\right)\right),
\]
where, in natural notations, $F_{Z,1}(z)(x):= \operatorname{pr}_E
F_Z(x, g(x), g'(x))$ and $F_{Z,2}(z)(x):= \operatorname{pr}_G
F_Z(x, g(x), g'(x))$. Second, we define a measure $\nu_Z:=
\left(F_{Z.1}(z) \right)_* \nu$; then
we define $F_{g, \nu}(z)$ by $F_{g, \nu}(z):=
L\left(\cdot , g_Z(\cdot ),g_Z'(\cdot )\right)\nu_Z$
(so, $g_0=g$, $\nu_0 =\nu$, and $F_{g, \nu}(0) =
L\left(\cdot , g(\cdot ),g'(\cdot )\right)\nu$).

For $x \in E$, $\Delta \in Z$, let
\begin{align*}
h_{1, \Delta}(x)&:= \left((F_{Z,1})'(0) \Delta\right) (x, g(x), g'(x)), \\
h_{2, \Delta}(x)&:=\left((F_{Z,2})'(0) \Delta\right) (x, g(x), g'(x)), \\
h_{3, \Delta}(x)&: = h_{2, \Delta}'(x).
\end{align*}
\begin{thm}
\label{thm1}
 The logarithmic derivative $\beta_{F_{g, \nu}}$ of $F_{g, \nu}$ at $0$ along $\Delta \in Z$ is given by
\begin{align*}
\left(\beta_{F_{g, \nu}}(0)\Delta\right)(x)&=
L_1'(x, g(x), g'(x))h_{1, \Delta}(x) +
L_2'(x, g(x), g'(x))h_{2, \Delta}(x) \\
& \quad + L_3'(x, g(x), g'(x))h_{2, \Delta}'(x) +
L\left(x, g(x),g'(x)\right) \operatorname{tr}h_{3, \Delta}(x) \\
& \quad  + L\left(x, g(x),g'(x)\right)
\beta^\nu\left(h_{2, \Delta}(x), x \right).
\end{align*}
\end{thm}

The theorem follows from the Leibniz rule for differentiation of the
product of a function and a measure, the chain rule, and formula
\eqref{rem_formula}.

\begin{rem}
The completely similar calculation can be applied when $\nu$ is not a
measure, but a pseudomeasure; hence an analog of Theorem \ref{thm1} is also
available for pseudomeasures.
\hfill $\lozenge$
\end{rem}

\begin{rem}
If $\nu$ is a translation invariant measure (in this case the dimension of
$E$ has to be finite) or pseudomeasure, then $\beta^\nu(h_{2,\Delta}(x),x)
= 0$ for any $x \in E$.
\hfill $\lozenge$
\end{rem}

\begin{thm}
\label{thm2}
{\rm(}``Noether Theorem''{\rm)} if $F_Z'(0)=0$, then for any $\Delta \in Z$,
we have
\begin{align*}
&L_1'(x, g(x), g'(x))h_{1, \Delta}(x) +
L_2'(x, g(x), g'(x))h_{2, \Delta}(x)
+ L_3'(x, g(x), g'(x))h_{2, \Delta}'(x)\\
& \qquad + L \operatorname{tr}h_{3, \Delta}(x)
+ L\beta^\nu\left(h_{2, \Delta}(x), x \right) = 0.
\end{align*}
\end{thm}
\noindent This follows from Theorem \ref{thm1}, noting that $F_Z'(0)=0$
implies $\left(\beta_{F_{g, \nu}}(0)\Delta\right)(x) = 0$ for all $x \in E$.

\begin{rem}
Theorem \ref{thm2} contains the first (direct) assertions of both theorems
usually called  the First and Second Noether Theorems. We quote them below
from the English translation in the excellent historical book
\cite{Ko2011} and keep Noether's original notations. Noether considers
functions $u_1(x), \ldots, u_\mu(x)$ for $x \in \mathbb{R}^n$. Then
she uses an invertible transformation of $x \in \mathbb{R}^n$, calls the
resulting variable $y \in\mathbb{R}^n$, and the transformed functions
$v_1(y), \ldots,v_\mu(y)$. She says that the transformations are part
of a group, but she never uses the group structure. Then she says
that ``an integral $I$ is an invariant of the group if it satisfies the relation
\[
\int\cdots \int f \left(x, u, \frac{\partial u}{\partial x},
\frac{\partial^2 u}{ \partial x} , \ldots  \right)dx =
\int\cdots \int f \left(y, v, \frac{\partial v}{\partial y},
\frac{\partial^2 v}{ \partial y} , \ldots  \right)dx
\]
integrated over an \emph{arbitrary} real domain in $x$, and over the
corresponding domain in $y$.'' The First Noether Theorem: ``If the
integral $I$ is invariant under a [group] $\mathfrak{G}_\rho$
\footnote{i.e., a family of invertible transformations depending on
$\rho$ parameters, not necessarily a group}, then
there are $\rho$ linearly independent combinations among the Lagrangian
expressions which become divergences---and conversely, that implies the
invariance of $I$ under a [group] $\mathfrak{G}_\rho$. The theorem
remains valid in the limiting case of an infinite number of parameters.''
The Second Noether Theorem: ``If the integral $I$ is invariant under
a [group] $\mathfrak{G}_{\infty\rho}$ depending on arbitrary functions
and their derivatives up to order $\sigma$, then there are $\rho$
identities among the Lagrangian expressions and their derivatives up
to order $\sigma$. Here as well the converse is valid.''
\hfill $\lozenge$
\end{rem}

\begin{rem}
The assumption above about arbitrariness of the domain of integration
means that actually one needs to study the integrands (but not the
integrals), which are products of a function and of a measure.
\hfill $\lozenge$
\end{rem}

\begin{rem}
In Theorems \ref{thm1} and \ref{thm2}, one considers the product of a
function and of a measure. Instead, one could substitute the measure $\nu$
by a product of a generalized density of $\nu$ and of the (translation invariant) Lebesgue pseudomeasure. By definition, the generalized
density has the same logarithmic derivative and hence, as far as the
formal calculations are concerned, nothing changes in carrying out
this more general case. On the other hand, the integral with respect to
the Lebesgue pseudomeasure of the generalized density is defined as a
limit of some finite-dimensional integrals and it is necessary to
ensure that the Leibniz rule holds; but this follows from the fact that
the Leibniz rule is valid for products of functions with the measure
$\nu$, because the finite-dimensional integrals in both cases are the same.
\hfill $\lozenge$
\end{rem}

\section{Quantum anomalies}
\label{sec:QA}
The calculations of the preceding section, where we integrated over
infinite dimensional spaces, are well adapted to problems of quantum
anomalies,  where one needs to integrate over an infinite dimensional
space of functions, the integrand
being the complex exponential of the action times either the Lebesgue
pseudomeasure or the Feynman pseudomeasure and a function depending on
initial data. If the derivative of the action with respect to a parameter,
on which the transformations of the domain of the action depend, is equal
to zero, then we can apply Theorem \ref{thm2}.

Let $\mathscr{E}$ be the phase space of a classical Hamiltonian system,
$\mathscr{E}:= Q \times P$, $Q$ and $P$ finite dimensional vector spaces,
$\mathfrak{h}: \mathscr{E} \rightarrow \mathbb{R}$ the Hamiltonian
function, and $\widehat{\mathfrak{h}}$
the pseudodifferential operator on $\mathcal{L}^2(Q)$ whose Weyl symbol
is $\mathfrak{h}$. Let $C([0,t], \mathscr{E})$ be the space of some
functions defined on $[0,t]$ taking values in $\mathscr{E}$; the
elements of $C([0,t], \mathscr{E})$ are pairs $\left(\xi_Q, \xi_P\right)$,
where $\xi_Q:[0,t] \rightarrow Q$, $\xi_P:[0,t] \rightarrow P$. Let
$f \in \mathcal{L}^2(Q)$ be the initial data for
the Cauchy problem for the Schr\"odinger equation ${\rm i}\psi'(t)
= \widehat{\mathfrak{h}} \psi(t)$ for a function $\psi: \mathbb{R}
\rightarrow \mathcal{L}^2(Q)$. Then the solution of the Cauchy problem for
this equation is given by the Feynman path integral with respect to the
Lebesgue pseudomeasure $\nu$, namely,
\[
\psi(t)(q) = \int_{C([0,t], \mathscr{E})} e^{{\rm i}\int_0^t \mathfrak{h}
\left(\xi_Q(\tau)+q, \xi_P(\tau)\right) d\tau}
e^{-{\rm i}\int_0^t \left\langle \dot{\xi}_Q(\tau),\, \xi_P(\tau)
\right\rangle d\tau} f\left(\xi_Q(t)+q\right)
\nu\left(d\xi_Q\, d\xi_P\right),
\]
where $\left\langle \cdot , \cdot \right\rangle: Q \times P\rightarrow
\mathbb{R}$ denotes the duality pairing. We do not discuss here the
definition of this integral (cf. \cite{SmSh2015}, where such an integral is not
explicitly defined). Instead of the integral with respect to the
pseudomeasure $\nu$, one can consider the product of the function
\[
\left(\xi_Q, \xi_P\right) \longmapsto  e^{-{\rm i}\int_0^t \left\langle
\dot{\xi}_Q(\tau), \,\xi_P(\tau)  \right\rangle d\tau}
\]
with $\nu$, which coincides with the Feynman pseudomeasure on the space
$C([0,t], \mathscr{E})$. To apply the technique of the preceding
section, we now make the identifications
\begin{itemize}
\item  $E$ is the space $C([0,t], \mathscr{E})$,
\item  $G$ is the space $\mathbb{C}$ of complex numbers,
\item $\mu(g, \nu)$ is the pseudomeasure which is the product of the function
\begin{equation*}
\label{eq2}
\left(\xi_Q, \xi_P\right) \longmapsto
e^{{\rm i}\int_0^t \mathfrak{h}
\left(\xi_Q(\tau)+q, \xi_P(\tau)\right) d\tau}
e^{-{\rm i}\int_0^t \left\langle \dot{\xi}_Q(\tau),\, \xi_P(\tau)
\right\rangle d\tau} f\left(\xi_Q(t)+q\right)
\end{equation*}
with the Lebesgue pseudomeasure $\nu$,
\item $F_Z$, a formally much simpler map $\Phi$ from an auxiliary
space $Z$ into the transformations of $C([0,t], \mathscr{E})$.
\end{itemize}

Then we have the following consequence of Theorem \ref{thm2}.

\begin{cor}
If the action and initial data are invariant with respect to transformations
of $\Phi(Z) \subset C([0,t], \mathscr{E})$, then the logarithmic derivative
of the pseudomeasure-valued function on $Z$, which is the sum of the
logarithmic derivatives of \eqref{eq2} and the Lebesgue pseudomeasure,
is equal to the logarithmic derivative
of the Lebesgue pseudomeasure and hence need not vanish. This non-vanishing
derivative is just the quantum anomaly {\rm(}cf. {\rm\cite{GoRaSm2015})}.
\end{cor}

\noindent \underline{Acknowledgment}: O.G.S. thanks the School of
Mathematics of the Shanghai Jiao Tong University for the excellent
working conditions provided by  during his visit in July 2016, when this
paper was written.

\end{document}